\documentclass[%
 aip,
 amsmath,amssymb,
 reprint,
]{revtex4-1}

\usepackage{graphicx}
\usepackage{dcolumn}
\usepackage{bm}

\usepackage[utf8]{inputenc}
\usepackage[T1]{fontenc}
\usepackage{mathptmx}
\usepackage{etoolbox}

\usepackage{xcolor}
\usepackage{soul}

\makeatletter
\def\@email#1#2{%
 \endgroup
 \patchcmd{\titleblock@produce}
  {\frontmatter@RRAPformat}
  {\frontmatter@RRAPformat{\produce@RRAP{*#1\href{mailto:#2}{#2}}}\frontmatter@RRAPformat}
  {}{}
}%
\makeatother
\begin{document}

\preprint{AIP/JCP}

\title[]{A statistical analysis of the first stages of freezing and melting of Lennard-Jones particles: Number and size distributions of transient nuclei}
\author{Patrice Porion}
 \email{patrice.porion@cnrs-orleans.fr}
\affiliation{ 
ICMN, CNRS, Universit\'e d'Orl\'eans, 1b rue de la F\'erollerie, CS 40059, 45071 Orl\'eans cedex 02, France }

\author{Jo\"el Puibasset}
 \email{joel.puibasset@cnrs-orleans.fr}
\affiliation{ 
ICMN, CNRS, Universit\'e d'Orl\'eans, 1b rue de la F\'erollerie, CS 40059, 45071 Orl\'eans cedex 02, France }

\date{\today}

\begin{abstract}
The freezing/melting transition is at the heart of many natural and industrial processes.
In the classical picture, the transition proceeds via the nucleation of the new phase, which has to overcome a barrier associated to the free energy cost of the growing nucleus.
The total nucleation rate is also influenced by a kinetic factor which somehow depends on the number of attempts to create a nucleus, that translates into a significant density of proto-nuclei in the system.   
These transient tiny nuclei are not accessible to experiments, but they can be observed in molecular simulations, and their number and size distributions can be acquired and analysed.
The number distributions are carefully characterized as a function of the system size, showing the expected behavior, with limited spurious effects due to the finite simulation box.
It is also shown that the proto-nuclei do exist even in the stable phase, in agreement with the fact that the (unfavorable) volume contribution to their free energy is negligible in the first stages of nucleation. Moreover, the number and size distributions evolve continuously between the stable and the metastable phases, in particular when crossing the coexistence temperature.
The size distributions associated to \textit{any} nucleus and to the \textit{largest} one have also been calculated, and their relationship recently established for bubbles in a liquid [J. Puibasset, J. Chem. Phys. 157, 191102 (2022)] has been shown to apply here. 
This is an important relation for free energy barrier calculations with biased molecular simulations.
\end{abstract}

\maketitle

Corresponding author: joel.puibasset@cnrs-orleans.fr  

\section{Introduction}
The freezing/melting transition is very common, but still not fully understood. The main difference with the liquid/vapor transition is the fact that it is qualitative (order/disorder transition) instead of quantitative (high vs low fluid density). 
This transition is of primary importance to understand different phenomena in nature and industry (\textit{e.g.} cloud formation in climate modeling, lubrication, tribology, cryopreservation or reactivity)\cite{RN3052, RN3050, RN3051} but also regarding fundamental concepts on order/disorder transitions (\textit{e.g.} dynamics of the transition, effect of confinement or premelting of surfaces).\cite{RN3024, RN3021,RN806}

Molecular simulation is a powerful tool to study the freezing/melting transition, in particular the molecular scale mechanisms underlying the transition.\cite{RN1255, RN3089, RN2917, RN3061, RN3098}
While in experiments melting always occurs at equilibrium by propagation from the outer surface where premelting occurs, in simulations, the introduction of periodic boundary conditions prevents premelting.
Therefore, it is possible to numerically monitor the experimentally inaccessible metastable (superheated) crystal, enabling to study the underlying mechanisms of bulk crystal destabilization during melting.\cite{RN3023,RN3021}

Molecular simulation is also a powerful tool to study the first stages of nucleation mechanism, in particular the activated nucleation of the new phase, inaccessible to experiments.
The quantity of interest is the probability that a nucleus reaches the size $s$. For subcritical sizes, this probability is also given by the equilibrium distribution of nuclei sizes in the system. This probability, denoted $p_a(s)$, for \textit{any} nucleus, gives access to the free energy profile $W(s)$ of the embryos through:\cite{RN2859,RN2913}

\begin{equation}
 W(s) = -kT \ln p_a(s) \label{eq_boltz}
\end{equation}
where $k$ is Boltzmann's constant and $T$ is the temperature.
This profile also gives access to the free energy barrier $W(s_c)$ entering the nucleation rate as given by the classical nucleation theory (CNT), where $s_c$ is the size of the critical nucleus.\cite{RN548,RN512,RN2920} 

Although $p_a(s)$ is in principle measurable in equilibrium molecular simulations, in practice, the simulations have to be biased to sample the regions where the free energy profile exceeds several $kT$.
The bias consists in monitoring the nucleus size from zero up to the critical size $s_c$ corresponding to the maximum of $W(s)$.
This requires to define an order parameter, combined with the umbrella sampling scheme of Torrie and Valleau.\cite{RN2885}
In a seminal work,\cite{RN2901,RN2909} it was proposed to choose the size of the \textit{largest} nucleus as the order parameter (it is unfortunately impossible to choose the size of \textit{any} nucleus). 
The corresponding probability distribution will be denoted $p_l(s)$. 
For large nuclei, the obtained distribution should correspond to the desired $p_a(s)$ since the probability to have more than one large nucleus in the simulation box is negligible.\cite{RN2901,RN2909}
However, for small sizes, one expects large differences between $p_l(s)$ and $p_a(s)$.
This can be seen by considering the small $s$ limit: $p_a(s)$ is expected to become large (since the free energy cost to produce small nuclei is negligible), while $p_l(s)$ is expected to become negligible (since $p_l(0)$ also corresponds to the probability to have \textit{no} nucleus in the phase, which is highly improbable).
Furthermore, for an homogeneous fluid, $p_a(s)$ is independent of the system size, which is consistent with Eq~\ref{eq_boltz}, while $p_l(s)$, being a maximum of a sample, strongly depends on the total number of nuclei in the system, and therefore depends on the system size.\cite{RN3049}
As recognized later, using $p_l(s)$ instead of $p_a(s)$ in Eq~\ref{eq_boltz} may lead to misinterpretations. \cite{RN2905, RN2893, RN2895, RN2907, RN2903, RN2847, RN2801}

To circumvent the problem, it was proposed to measure directly the exact $p_a(s)$ for the small values of $s$ where the disagreement is strongest, and to complete the distribution for large $s$ values by using $p_l(s)$ corrected by a multiplicative constant to ensure continuity. 
This approach has however some limitations, and introduces an approximation that cannot be quantified. \cite{RN2899}
Very recently, in the context of bubble nucleation in a metastable liquid, a general relation was proposed that relates $p_a(s)$ and $p_l(s)$:\cite{RN2954}
\begin{equation}
 p_a(s) = \frac{p_l(s)}{\lambda_0 \Pi_l(s)} \label{eq_pl_to_pa_conti}
\end{equation}
where $\Pi_l(s)$ is the cumulative distribution function of $p_l(s)$ and $\lambda_0$ is a constant.
This approach presents the advantage that it requires only one biased simulation series to measure accurately $p_l(s)$ from zero up to the critical size, which is then converted straightforwardly into $p_a(s)$. 

The objective of this paper is to explore the robustness of this relation in the context of the solid-liquid transition.
In particular, one has to check that the distribution $\phi(n)$ of the number $n$ of spontaneous nuclei in the stable and metastable phases follows the Poisson law, leading to the Eq.~\ref{eq_pl_to_pa_conti}, where $\lambda_0$ is the associated Poisson parameter.
A direct calculation of $p_a(s)$ and $p_l(s)$ is then performed is order to validate the relation.
Furthermore, since the solid can frequently present different stable or metastable structures (FCC, HCP,  etc.), it is possible to explore different transitions starting from all possible solid structures. 

The paper is divided as follows: the numerical model is first presented, including the calculation of important parameters characterizing the structure and thermodynamics of the fluid. The distributions of the number of nuclei $\phi(n)$ are then established for different phases, and the Fourier formalism is used to characterize the distributions independently of the system size. Their evolution with temperature is also established. The paper then focuses on the nuclei size distributions $p_a(s)$ and $p_l(s)$, with emphasis on their differences, and the verification of Eq.~\ref{eq_pl_to_pa_conti} that relates $p_l(s)$ to $p_a(s)$. Finally, the free energy profiles are calculated to discuss the very first stages of nucleation.
 
\section{Methods}
\subsection{Model}
It is well-known that the interactions in solid phases are in most cases not well approximated by pair-wise additive potentials. For example, this is the case for metals, where many-body potentials are better approximations. 
It is however assumed that the generic properties we are interested in here are mainly controlled by thermal activation, with a free energy profile comprising essentially volume and surface contributions.
We therefore consider a simple atomic model based on the truncated and linearly shifted (12-6) Lennard-Jones potential (continuous forces). The energy parameter and the kinetic diameter are denoted $\epsilon$ and $\sigma$, and the cutoff distance is taken equal to 3$\sigma$.  From now on, the quantities are expressed in reduced units: $\sigma$ for distances, $\epsilon$ for energies, $\epsilon/k$ for temperatures, $\epsilon/\sigma^3$ for pressures, etc. 

The molecular dynamics (MD) simulations are done in the isothermal-isobaric NPT ensemble using the Large-scale Atomic/Molecular Massively Parallel Simulator (LAMMPS)\cite{RN3018} with Nos\'e/Hoover thermostat and barostat.\cite{RN51, RN3029, RN3030, RN3035} The time increment is taken equal to $10^{-3} \sqrt{m\sigma^2/\epsilon}$ and configurations are sampled every 1000 steps for statistical analysis. Different system sizes are considered, ranging from $N=$ 216 to 16384 atoms. For pure phases simulations, the initial boxes are perfectly cubic; for two-phases simulations, the box is twice as long in one direction (juxtaposition of two cubic boxes). During the NPT runs, the three box dimensions may either be left free to fluctuate independently or can be coupled: the results are the same. 

The runs are performed at different temperatures, in the stable and the metastable regions of the phase diagram. The strongly metastable region is discarded, so that the transitions are never observed during the course of the simulations. The system is thus in local equilibrium, and the results are independent of the way the thermodynamic conditions are reached. In practice, the temperature changes are done at the rate of 0.1 in reduced units, followed by thermalization during 10$^6$ time steps before data acquisition.

\subsection{Freezing, melting and coexistence temperatures}
The phase diagram for the Lennard-Jones fluid shows a triple point slightly below $T=0.7$, with a stable FCC solid phase.\cite{RN3033, RN3031, RN2264, RN3019, RN3054}  The HCP structure is (slightly) metastable at that temperature, and becomes stable below $T=0.4$. On the other hand, the BCC structure is always unstable. The exact position of coexistence lines depends on the truncation used to calculate the interactions and the possible long range corrections, but the phase diagram looks qualitatively the same.   

\begin{figure}[t]
\includegraphics[width=1.0\columnwidth]{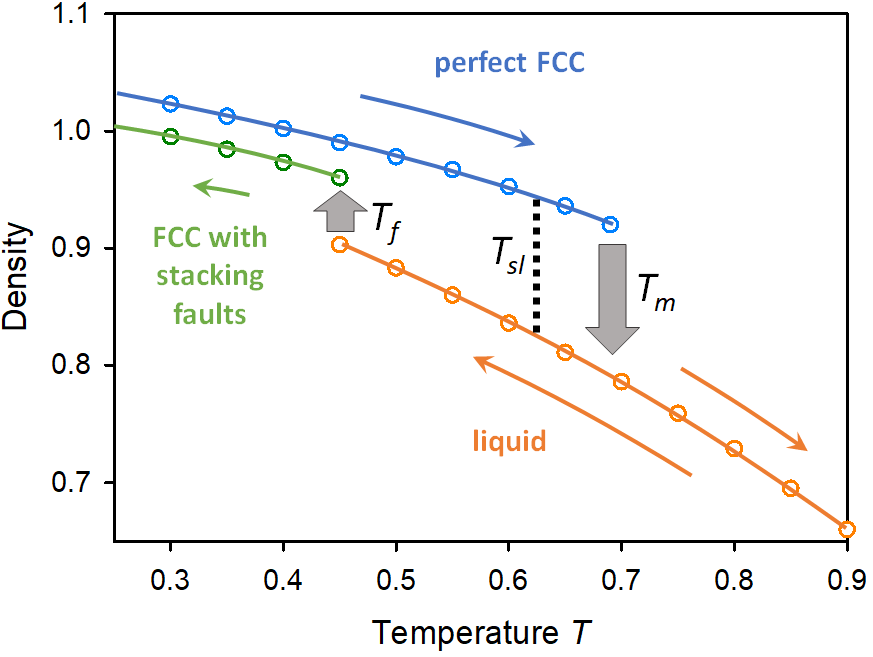}
\caption{\label{fig_hysteresis} Evolution of the density of a Lennard-Jones system (500 atoms) at $P=0$ during an ascending temperature ramp from $T=0.3$ to 0.9, followed by a descending ramp down to 0.3 (thin arrows). The large vertical arrows correspond to the irreversible transitions between (i) the initial perfect FCC crystal (upper blue branch) and the liquid (lowest orange branch), and, (ii) between the liquid and the imperfectly crystallized solid (intermediate green branch). These transitions define the melting ($T_m$) and freezing ($T_{f}$) temperatures respectively. The solid-liquid equilibrium occurs in between, at $T_{sl}$ (see text).      
}
\end{figure}

The present work focuses on the solid-liquid transition at low pressure, i.e. in the vicinity of the triple point (lower bound of the solid-liquid coexistence line).
These dense phases being cohesive, they remain metastable below that point, making it possible to observe solid-liquid coexistence well below the triple point pressure (down to possibly negative values).
In our case, the triple point pressure being quite small (around $10^{-3}$ in reduced units for the full Lennard-Jones), we work at zero pressure, without introducing measurable differences compared to the triple point.
In particular, the density and the structure (FCC, HCP or the radial distribution function for the liquid) are essentially unaffected by the small change in pressure. For instance, the relative variation of the density is only $4\times 10^{-5}$ between the triple point and zero pressure; for comparison, the relative density variation associated with the melting of one single atom at $T=0.62$ in a FCC crystal comprising 500 atoms is $\sim 3\times 10^{-4}$, i.e. seven times larger (see Fig.~\ref{fig_hysteresis}).
The only noticeable difference is the fact that, strictly speaking, the liquid and solid phases at zero pressure are both metastable with respect to the infinitely diluted phase (i.e. the gas phase at zero pressure). 
However, to simplify the discussion, the metastability with respect to the gas phase will be ignored, and the liquid or solid phase with the lowest free energy will be qualified as stable. 

As previously mentioned, the truncation of the potential slightly influences the phase diagram, in particular the solid-liquid coexistence line. The coexistence temperature at zero pressure thus needs to be evaluated.  
We focus on the transition between the liquid and the FCC crystal which is the most stable structure.  
Starting from a FCC crystal comprising 500 atoms with periodic boundary conditions, MD simulations with increasing temperature shows a melting transition around $T_m=0.69$ (Fig.~\ref{fig_hysteresis}).
Cooling back the liquid, one observes freezing around $T_f = 0.45$, revealing a large freezing/melting hysteresis. Furthermore, one observes imperfect crystallization upon cooling, with a significant number of defects in the FCC structure resulting in a lower density and higher energy (see Fig.~\ref{fig_hysteresis}).
It is emphasized that $T_m$ and $T_f$ are only approximately positioned since they depend on the initial conditions and the length of the MD run as well as the system size (stochastic transitions).  

This freezing/melting hysteresis defines the largest possible temperature range where it is possible to observe both solid and liquid phases. In the following, the calculations will be done in a slightly smaller interval, between $T = 0.46$ and 0.68, to avoid spurious effects close to the transitions.

\begin{figure}[t]
\includegraphics[width=1.0\columnwidth]{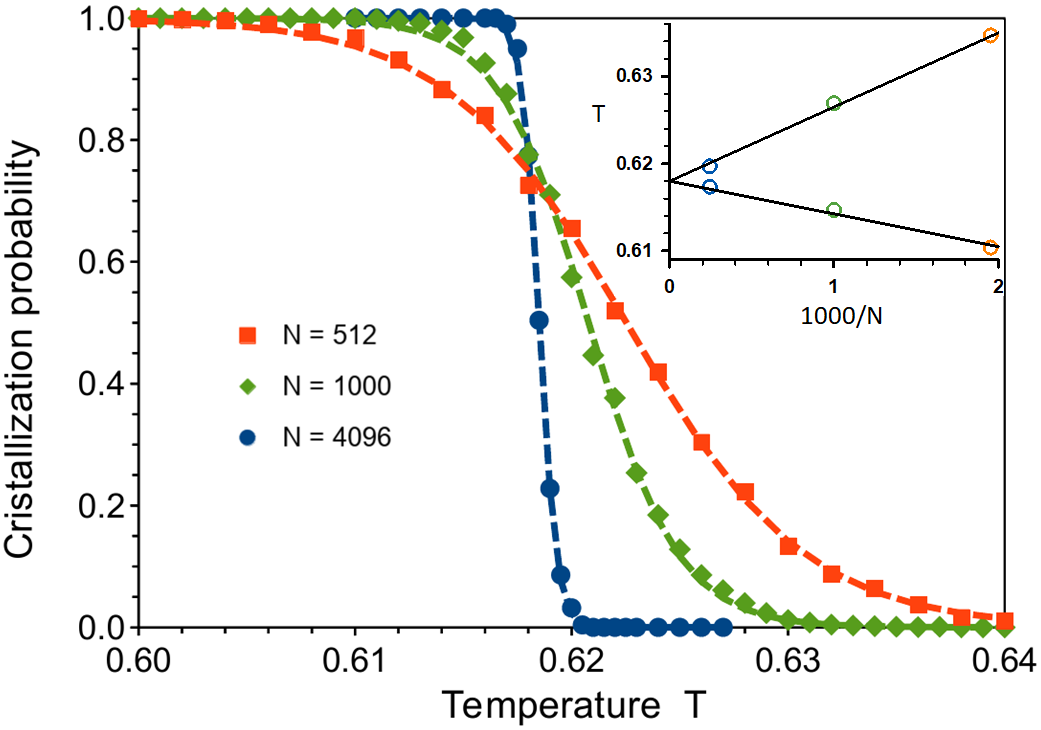}
\caption{\label{fig_T_coex} Symbols: crystallization probability for a system initially comprising an FCC crystal in contact with an equal number of liquid atoms, as a function of the temperature, at $P=0$. Three system sizes are considered (total number of crystal+liquid atoms $N$= 512, 1000 and 4096). The crystallization probability is calculated based on a large number of initial conditions: 2000 runs for $N$ = 512 and $N$ = 1000; 500 runs for $N$ = 4096. Lines: sigmoid fits. Inset: lower and upper bound for the coexistence temperature as given by a crystallization probability between 5 and 95\%.
}
\end{figure}

The temperature hysteresis is split into two sub-intervals by the solid-liquid equilibrium temperature $T_{sl}$: below $T_{sl}$, the solid is stable while the liquid is metastable; above $T_{sl}$ it is the opposite (see Fig.~\ref{fig_hysteresis}). 
The solid-liquid equilibrium has been determined by means of the two-phase method.\cite{RN3033} A cubic piece of FCC crystal comprising $N/2$ atoms is put in contact with a liquid box with the same number of atoms, producing a two-phase system comprising $N$ atoms.
The liquid/solid interface is first rapidly equilibrated with the FCC atoms kept frozen.
MD runs are then performed for the $N$ atoms in the isothermal-isobaric ensemble at $P$ = 0 at different temperatures.
In each case, the evolution of the total number of FCC atoms is monitored. At low $T$, the liquid region invariably fully crystallizes, while at high $T$ the crystal always melts. 
For intermediate temperatures, the system may either crystallize or melt depending on the initial conditions. One can define a probability towards crystallization by monitoring different initial conditions and counting the number of configurations that lead to a crystal. 
Figure~\ref{fig_T_coex} shows the evolution of this probability with the temperature $T$ for three system sizes. 
As can be seen, the crystallization probability decreases from one, at low temperature, down to zero, at high $T$,  following a sigmoid curve. As expected, the larger the system, the sharper the transition. The solid-liquid equilibrium temperature $T_{sl}$ can be defined as the temperature where the crystallization probability equals 0.5 (crystallization and melting are equiprobable), and an interval can be defined by imposing a crystallization probability between 5 and 95\% (see inset).
As can be seen, $T_{sl}$ slightly varies with the system size, with values around 0.62, in agreement with previous determinations based on other techniques.\cite{RN3041, RN3036, RN3043} 

\subsection{Order parameter}
The local lattice structure around an atom is determined using the Polyhedral Template Matching (PTM) method.\cite{RN3022}
For each atom of the sample, its local structure is compared to that of a list of templates (FCC, HCP, BCC, etc.), where ``local'' means that one includes the neighbors within the first minimum of the radial distribution function at temperature $T$.
For each atom \textit{and} template, the matching level is quantified by a scalar (denoted RMSD for root-mean-square deviation).
Each atom is then assigned the lowest RMSD value among all templates, and temporarily tagged with the template that matches the best with its local structure, i.e. the one giving the lowest RMSD. 
By construction, the RMSD is a scalar ranging from zero to infinity.

\begin{figure}[t]
\includegraphics[width=1.0\columnwidth]{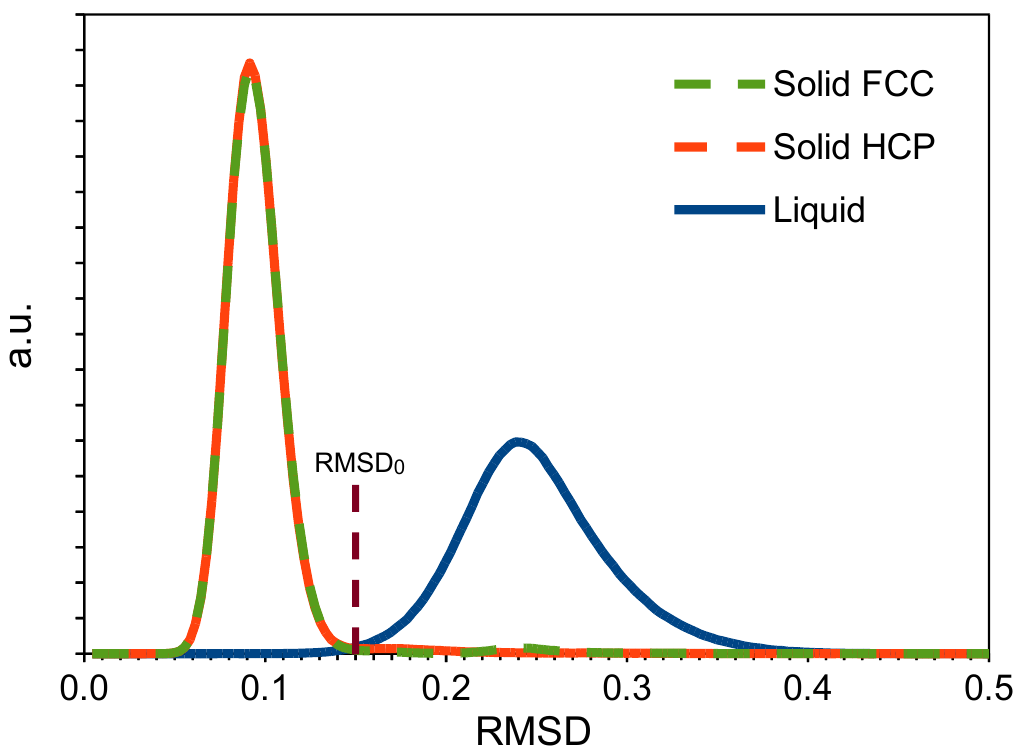}
\caption{\label{fig_rmsd} RMSD distributions calculated for the three samples corresponding to the phases of interest for this work: a FCC crystal (500 atoms), an HCP crystal (512 atoms) and a liquid sample (500 atoms), at $P$ = 0 and $T$ = 0.62. The optimal value of the RMSD cutoff (RMSD$_0$) at the intersection between the liquid and the two solid curves, is given by the vertical dashed line at RMSD$_0$ = 0.15.
}
\end{figure}

Figure~\ref{fig_rmsd} gives the RMSD distributions calculated for the three phases of interest for this work (an FCC crystal with 500 atoms, an HCP crystal with 512 atoms, and a liquid sample with 500 atoms) at $P$ = 0 and $T$ = 0.62, corresponding to the FCC-liquid coexistence temperature previously determined.
The RMSD values for the two crystalline phases (FCC and HCP) are identically distributed around the nominal value 0.09 with similar widths, each phase matching best with its corresponding template (FCC or HCP), as expected.
On the other hand, the RMSD density for the liquid phase is distributed around a larger value (0.24) with a larger spreading, and the corresponding RMSD values correspond to FCC, HCP or BCC templates, in proportion corresponding approximately to 10, 70 and 20\% respectively.

As can be seen, the RMSD distributions for the solid phases (FCC or HCP) and for the liquid phase are well separated. The intersection, denoted RMSD$_0$, is used as a threshold to identify the local structure of the atoms in \textit{any} given molecular configuration, as follows. 
Each atom is assigned the lowest RMSD value among all templates. 
If its RMSD falls below RMSD$_0$, the atom is tagged with the crystalline structure according to the corresponding RMSD optimal matching. On the other hand, if the RMSD is larger than RMSD$_0$, the local structure is identified as disordered and the atom is tagged as liquid.
In agreement with other authors, we found RMSD$_0$ = 0.15 (see Fig.~\ref{fig_rmsd}).\cite{RN3045, RN3046}
The PTM method is implemented in the Ovito software used to analyse the configurations.\cite{RN3044}

\subsection{Nuclei: identification and statistics}
\begin{figure}[b]
\includegraphics[width=1.0\columnwidth]{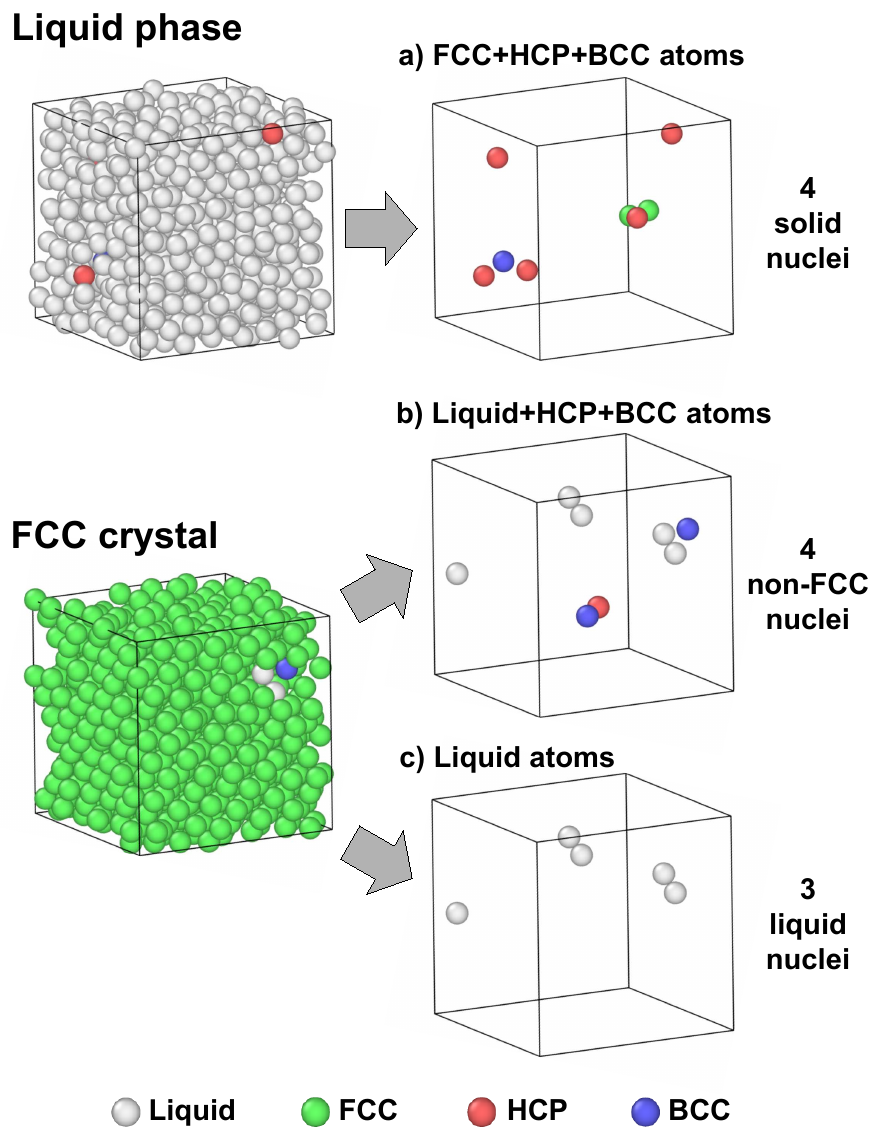}
\caption{\label{fig_clusterisation} Snapshots (generated with Ovito software\cite{RN3044}) illustrating the procedure to detect the solid nuclei in the liquid (a), the defective (non-FCC) nuclei in the FCC crystal (b), and the pure liquid nuclei in the FCC crystal (c). Left column: all atoms are shown; right column: only the atoms corresponding to the selected structures are shown, the nuclei being defined as clusters of these atoms. The color code corresponds to the local structure as given by the PTM method. 
}
\end{figure}

Let us first consider nucleation in the liquid phase. Visual inspection shows that solid-like atoms of different structures (FCC, HCP or BCC) appear spontaneously in the system [see Fig.~\ref{fig_clusterisation}(a)].
Quite remarkably, the relative abundance is in favor of HCP, at odds with the fact that FCC is the most stable structure for the bulk solid (at the coexistence temperature $T$ = 0.62, the HCP-to-FCC ratio equals 3.7).
This confirms that the relative stability of tiny crystallites in liquid does not necessarily follow the bulk behavior.\cite{RN2987, RN3095, RN2936}
We also observe that BCC defects are the less abundant. This possibly results from the instability of the BCC structure for the Lennard-Jones potential, and does not support the proposed idea that BCC could be favored in crystallization as being an intermediate structure between liquid and solid.\cite{RN30702, RN307042, RN302714, RN30132, RN30105, RN3091}
Furthermore, these solid-like atoms, which are mostly isolated, sometimes group into clusters that are not necessarily pure FCC or HCP, as can be seen on the example shown in Fig.~\ref{fig_clusterisation}(a).
We are therefore led to define the clusters without specifying the particular solid structure. The nuclei are thus defined as clusters of any solid atoms (FCC, HCP and BCC), the cluster grouping criterion being chosen to be the largest value, among the different crystal structures, of the first minimum of the radial distribution function at each temperature $T$. 

Let us now consider the tiny nuclei appearing in the FCC crystal. Visual inspection shows that liquid-like atoms are not the only defects that appear within the FCC crystal: HCP or BCC atoms can also be observed [see Fig.~\ref{fig_clusterisation}(b)]. As shown, different defective (non-FCC) atoms can appear in the same cluster: we are therefore led to define the nuclei in the FCC crystal as clusters of non-FCC atoms (including liquid, HCP and BCC), by analogy with what we did for the solid nuclei in the liquid phase. Note that in this case, the distance criterion for clustering corresponds to the first minimum of the radial distribution function of the liquid (larger than that of the solid).

However, it is also possible to restrict the cluster analysis to liquid-like atoms: this is what is depicted in Fig.~\ref{fig_clusterisation}(c). In that case, only liquid-like atoms are taken into account to define the nuclei, with the corresponding cluster criterion. These clusters will be denoted ``liquid nuclei''.
It is not clear which definition of nuclei (liquid or non-FCC) is the best to describe nucleation in solids. Obviously, one expects measurable differences for small clusters comprising few atoms. For instance, Fig.~\ref{fig_clusterisation}(b) and \ref{fig_clusterisation}(c) illustrate a case where the number and the size of nuclei differ depending on the criterion. However, for large nuclei relevant to calculate nucleation barriers, their composition is expected to be dominated by the most stable phase, and therefore, both criteria should, at the end, give similar results.   

For the HCP crystal, the situation is very similar to the FCC case: the nuclei may either be defined as clusters of non-HCP atoms (i.e. liquid, FCC and BCC), or clusters of liquid-like atoms solely.

This procedure allows to detect the nuclei appearing in the system for each atomic configuration. We denote $n$ the total number of nuclei of all sizes for each configuration, and $\phi (n)$ the corresponding distribution over all configurations, normalized to one. The size $s$ of the nuclei is defined as the total number of atoms in each cluster. The size densities $p_a(s)$ (all nuclei sizes) and $p_l(s)$ (largest nuclei size) resume to discrete histograms since the nuclei sizes $s$ are defined as their number of atoms. Note that each configuration contributes once to $p_l(s)$ (exactly one largest nucleus per configuration, possibly of size zero), leading to a histogram normalized to one after division by the number of configurations. On the other hand, each configuration contributes as many times to $p_a(s)$ as the number of nuclei in the configuration. Therefore, after building the histogram over all configurations, it has to be divided by the total number of nuclei over all configurations, so that it is normalized to one.
The number of configurations used for the averages ranges from 20000 for the largest systems to $10^6$ for the smallest ones.
Note also that $p_a(s)$ is defined only for non-zero sizes, while $p_l(s)$ is defined for any size, including zero; in particular, $p_l(0)$ is the probability that the molecular configuration contains no nuclei.

\section{Results}
\subsection{Distribution of the number of nuclei $\phi(n)$}

\subsubsection{Influence of the system size}
\begin{figure}[b]
\includegraphics[width=1.0\columnwidth]{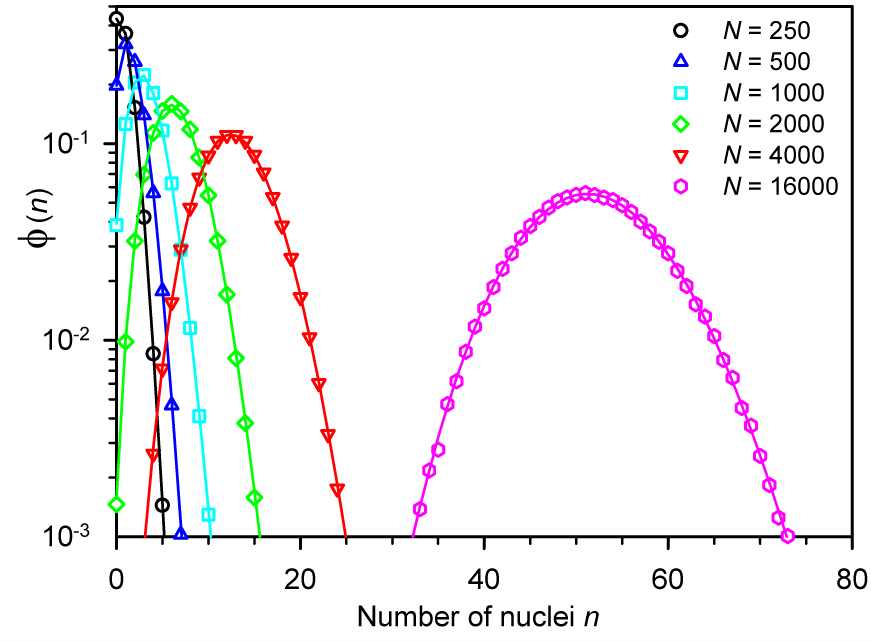}
\caption{\label{fig_distrib_phi_Liq_vs_size} Symbols: distributions $\phi_N(n)$ of the number $n$ of solid nuclei (FCC+HCP+BCC) in the liquid at $P=0$ and $T=0.62$, shown in log-linear scale and for different system sizes $N$ given in the figure. Lines are fits with Poisson distributions.
}
\end{figure}

Typical $\phi_N(n)$ (in this section and the next, the dependence with the system size $N$ is shown as a subscript) are given in Fig.~\ref{fig_distrib_phi_Liq_vs_size} in the case of solid nucleation in the liquid, at the solid-liquid coexistence temperature $T=0.62$ and for different system sizes ranging from $N$ = 250 to 16000 atoms. 
The distributions exhibit a maximum, corresponding to a typical number of nuclei, that increases linearly with the system size.
The width of the distributions also increases with the system size.
One expects that nucleation should follow a Poisson distribution, at least in the low rate regime where the nucleation events are spatially independent.\cite{RN2893, RN3047}
This is confirmed by the good quality of the Poisson fits shown in Fig.~\ref{fig_distrib_phi_Liq_vs_size}.

\begin{figure}[b]
\includegraphics[width=1.0\columnwidth]{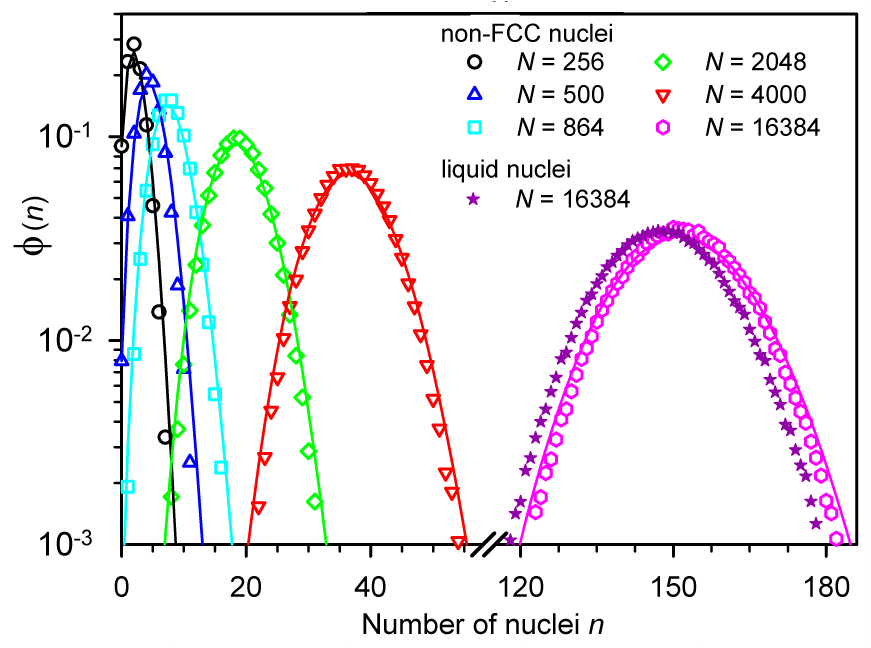}
\caption{\label{fig_distrib_phi_FCC_vs_size} Symbols: distributions $\phi_N(n)$ of the number $n$ of nuclei in the FCC crystal at $P=0$ and $T=0.62$, shown in log-linear scale and for different system sizes $N$ given in the figure. The empty symbols refer to the non-FCC nuclei, while the stars refer to the pure liquid nuclei (shown only for the largest system). Lines are fits of the non-FCC nuclei distributions with the Poisson law. 
}
\end{figure}

The symmetric situation, corresponding to the nucleation in the FCC crystal at the same temperature and pressure, gives similar results shown in Fig.~\ref{fig_distrib_phi_FCC_vs_size}. 
For comparison, the distributions for the largest system are calculated either with all non-FCC (liquid+HCP+BCC) atoms to define the clusters, or with only liquid atoms.
The main difference appears as a shift of the distribution to slightly lower numbers of nuclei for pure liquid clusters (see Fig.~\ref{fig_distrib_phi_FCC_vs_size}).
However, the relative difference is less than 3\%.
Furthermore, this difference is expected to decrease for larger nuclei (closer to the nucleation barrier). 
Focusing on non-FCC nuclei, the behavior with the system size is essentially the same as for solid nuclei in liquid, the main difference being the larger values for the average number of nuclei, showing that, at the same temperature and pressure, more nuclei are detected in the solid compared to the liquid.
The second important difference is that the Poisson fits exhibit small disagreements, mainly for the lowest values of the distributions ($\phi_N(n) \leq 10^{-2}$, see Fig.~\ref{fig_distrib_phi_FCC_vs_size}).

The same system size analysis has been done for the HCP crystal, showing distributions almost indistinguishable from those obtained in the FCC crystal for similar system sizes (for instance 500 and 512 or 4000 and 4096 in the FCC and HCP crystals respectively). We observe in particular the same departure from the Poisson law. In order to elucidate whether this non-Poisson behavior is due to finite size effects, we need to compare quantitatively the distributions associated to different system sizes, as follows.  

\subsubsection{Finite size analysis}
To characterize the evolution of the distributions with the system size $N$, it is useful to notice that the distribution for the system of size $N_1 + N_2$ is the convolution of the distributions for the sub-systems of size $N_1$ and $N_2$. 
For instance, the probability to observe no nuclei in the total system is $\phi_{N_1+N_2}(0)=\phi_{N_1}(0)\phi_{N_2}(0)$.
More generally, one has $\phi_{N_1+N_2}(n)=\sum_{i=0}^{n}\phi_{N_1}(i)\phi_{N_2}(n-i)$.

\begin{figure}[b]
\includegraphics[width=1.0\columnwidth]{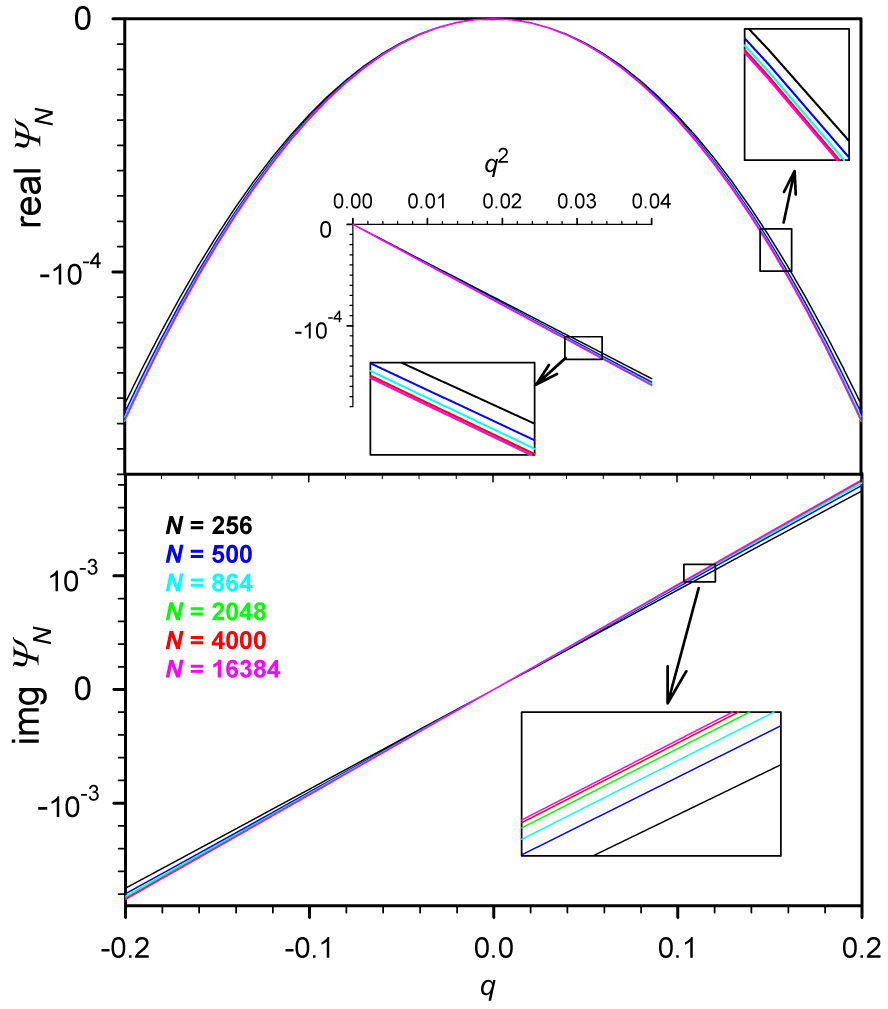}
\caption{\label{fig_lnTFdistrib_FCC} Real and imaginary parts of $\psi_N(q)$ (Eq~\ref{eq_psi}) for the nuclei distributions $\phi_N(n)$ (shown in Fig.~\ref{fig_distrib_phi_FCC_vs_size}) in the FCC crystal at $P=0$ and $T=0.62$ and for different system sizes $N$ given in the figure. Upper panel: real part versus $q$ (main) and $q^2$ (insert). Lower panel: imaginary part versus $q$. The small insets are local enlargements to evidence the weak system size dependence.
}
\end{figure}

\begin{figure}[b]
\includegraphics[width=1.0\columnwidth]{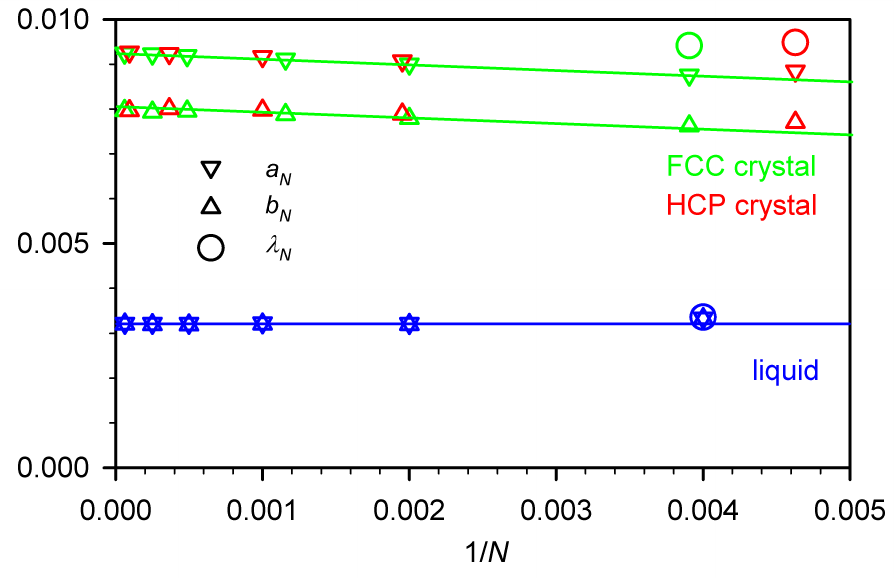}
\caption{\label{fig_a_b_vs_size} Linear ($a_N$, down triangles) and quadratic ($b_N$, up triangles) coefficients (Eq~\ref{eq_ab}) for, respectively, the imaginary and real parts of $\psi_N(q)$ (see Fig.~\ref{fig_lnTFdistrib_FCC}), as a function of $1/N$, for nucleation in the liquid (blue points) and in the FCC (green) and HCP (red) crystals. A best fit is shown with one (liquid) or two parallel (FCC crystal) lines. Circles: $\lambda_N = -N^{-1}\ln\phi_N(0)$ when defined with a good accuracy (i.e. for the smallest systems).
}
\end{figure}

Therefore, introducing the Fourier formalism for $\phi_N(n)$:
\begin{equation}
\tilde\phi_N(q) = \sum_{n=0}^{\infty} \phi_N(n)e^{iqn}           
\end{equation}
and
\begin{equation}
\phi_N(n) = \frac{1}{2\pi} \int_{-\pi}^{\pi} \tilde\phi_N(q)e^{-iqn} dq
\end{equation}
one has the remarkable property:
\begin{equation}
\tilde\phi_{N_1+N_2}(q)= \tilde\phi_{N_1}(q)\tilde\phi_{N_2}(q)
\end{equation}
meaning that $\ln \tilde\phi_{N}(q)$ is extensive with the system size $N$.
We therefore introduce the normalized quantity
\begin{equation}
\psi_N(q) = N^{-1} \ln \tilde\phi_{N}(q)  \label{eq_psi}
\end{equation}
to characterize and compare the distributions for different system sizes.
This quantity is shown in Fig.~\ref{fig_lnTFdistrib_FCC} in the case of nucleation in the FCC crystal at $P=0$ and $T=0.62$. 
The results for nucleation in HCP are essentially indistinguishable from FCC, and, for nucleation in liquid, they are qualitatively the same except that the system size dependence is not visible. For these reasons they are not shown. 
As evidenced in Fig.~\ref{fig_lnTFdistrib_FCC}, the imaginary part is essentially linear with the frequency $q$, while the real part is quadratic (see inset).
A best fit with the equation 
\begin{equation}
\psi_N(q) = ia_Nq -0.5b_Nq^2  \label{eq_ab}
\end{equation}
gives the coefficients $a_N$ and $b_N$ (corresponding to the first cumulants) as a function of the system size $N$.
As shown in Fig.~\ref{fig_a_b_vs_size}, $a_N$ and $b_N$ vary linearly with $1/N$, as expected in standard finite size analysis, suggesting the values $a_\infty=9.23\times 10^{-3}$ and $b_\infty=8.02\times 10^{-3}$ for nucleation in the infinite FCC crystal (quantitatively the same result is obtained for the HCP structure).
On the other hand, the coefficients $a_N$ and $b_N$ associated to nucleation in the liquid phase exhibit no measurable dependence with $N$ and are almost identical: $a_\infty=3.22\times 10^{-3}$ and $b_\infty=3.20\times 10^{-3}$.

For independent nucleations, one expects that $\phi_N$ should follow a Poisson distribution $\phi^P_N(n)=e^{-\Lambda_N} \Lambda_N^n/n!$, with a parameter $\Lambda_N$ (equal to the average number of nuclei in the system) proportional to the system size: $\Lambda_N = N\lambda$. 
In this case, 
\begin{eqnarray}
    \psi_N(q) = N^{-1} \ln\tilde\phi^P_{N}(q) = \lambda \left( e^{iq} -1 \right)   \nonumber \\    
=  i\lambda q - 0.5\lambda q^2 + O(q^3)
\end{eqnarray}
meaning that the coefficients $a_N=b_N=\lambda$.
This is what we observe for nucleation in the liquid phase whatever the size of the system (see Fig.~\ref{fig_a_b_vs_size}). The fact that in the FCC and the HCP crystals the two coefficients $a_N$ and $b_N$ slightly differ (by a constant independent of the system size as shown by the two parallel lines in Fig.~\ref{fig_a_b_vs_size}) suggests that the distributions of the number of nuclei slightly depart from the Poisson law even for the infinite system.
The disagreement is however moderate ($<10\%$ in the lowest part of the distributions where the disagreement is strongest, see Fig.~\ref{fig_distrib_phi_FCC_vs_size}).
It is not yet understood why the distributions depart from the Poisson statistics, in a way independent of the system size. This suggests that correlations between nuclei could explain this discrepancy. 

Note that for the Poisson law $\phi^P_N$, $\lambda = -N^{-1}\ln\phi^P_N(0)$. By analogy, for any $\phi_N$, we denote $\lambda_N = -N^{-1}\ln\phi_N(0)$, which now depends on the system size $N$. 
Figure~\ref{fig_a_b_vs_size} shows that $\lambda_N$ (circles) is an excellent guess for the coefficient $a_\infty$ of the infinite system, at least as long as the system is small enough so that $\phi_N(0)$ can be determined with a good accuracy.
It is however not yet understood why this quantity, that can be determined only for the smallest systems, is a good guess for $a_\infty$.

This finite size analysis shows that $N \geq 2000$ is a good compromise between computational cost and accuracy for all systems (less than $1\%$ discrepancy compared to the infinite system).
In the following, all calculations will adopt this prescription; the system size being imposed and explicitly given for each situation, the subscripts $N$ will be dropped.

\subsubsection{Influence of the temperature and the nature of the transition} 
\begin{figure}[]
\includegraphics[width=1.0\columnwidth]{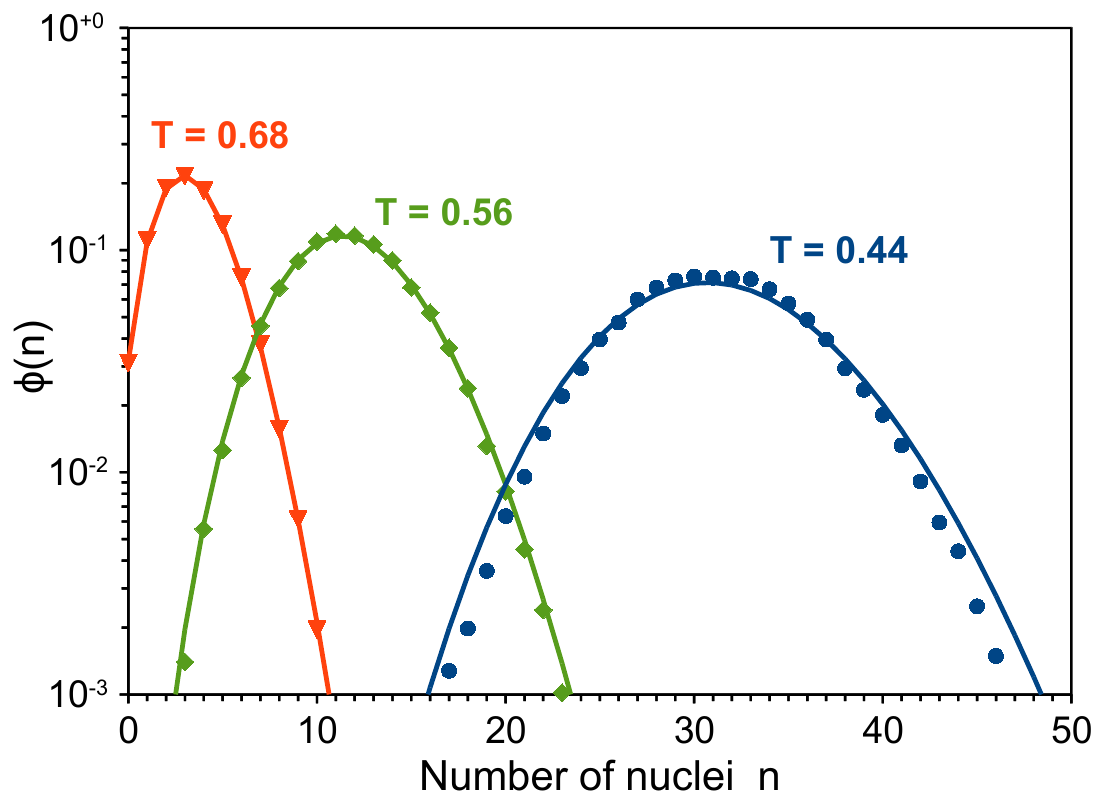}
\caption{\label{fig_distrib_phi_Liq_vs_T} Symbols: distributions $\phi(n)$ of the number of solid (FCC+HCP+BCC) nuclei in the liquid system with $N=2000$ at $P=0$ and for different temperatures given in the figure. The lines are fits with Poisson distributions. 
}
\end{figure}

The distributions $\phi(n)$ of the number of solid nuclei in the liquid are shown in Fig.~\ref{fig_distrib_phi_Liq_vs_T} for three temperatures (close to $T_f$, $T_m$ and in between).
As can be seen, the lower the temperature, the larger the probability to observe a large number of nuclei and the lower the probability to observe a small number of nuclei.
It is also visible that the Poisson fits are less accurate at low temperature, suggesting correlations between nuclei when their density increases.

\begin{figure}[]
\includegraphics[width=1.0\columnwidth]{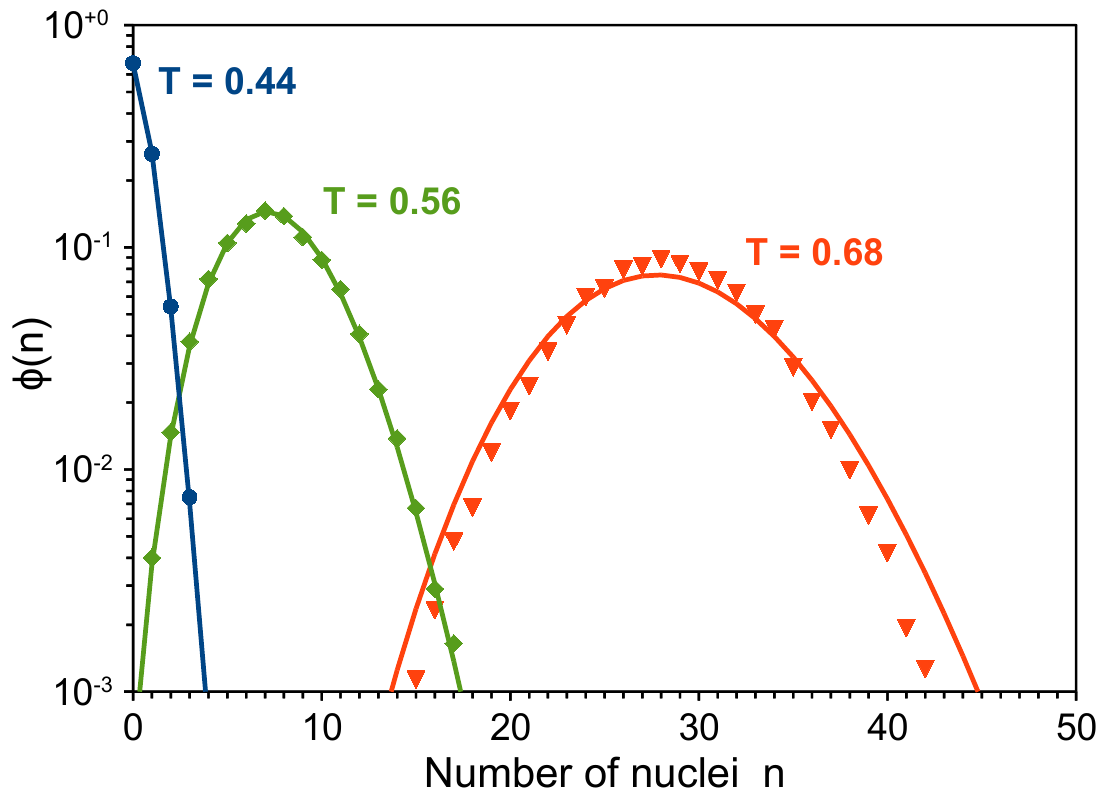}
\caption{\label{fig_distrib_phi_fcc_vs_T} Symbols: distributions $\phi(n)$ of the number of non-FCC (liquid+HCP+BCC) nuclei in the FCC crystal of size $N=2048$ at $P=0$ and for different temperatures given in the figure. The lines are fits with Poisson distributions.
}
\end{figure}

The inverse situation has been studied, corresponding to nucleation in the FCC crystal. The distributions of the number of non-FCC nuclei are given in Fig.~\ref{fig_distrib_phi_fcc_vs_T} for different temperatures.
The trend with temperature is now reversed, with the average number of nuclei increasing for larger $T$. The Poisson fits also exhibit good matching at low temperature, while the highest temperature clearly exhibits departure from the expected Poisson distribution for independent events. As previously, this correlates with the increase in the number of nucleation events. 
Quantitatively similar results are obtained for nucleation in the HCP crystal. 

\begin{figure}[]
\includegraphics[width=1.0\columnwidth]{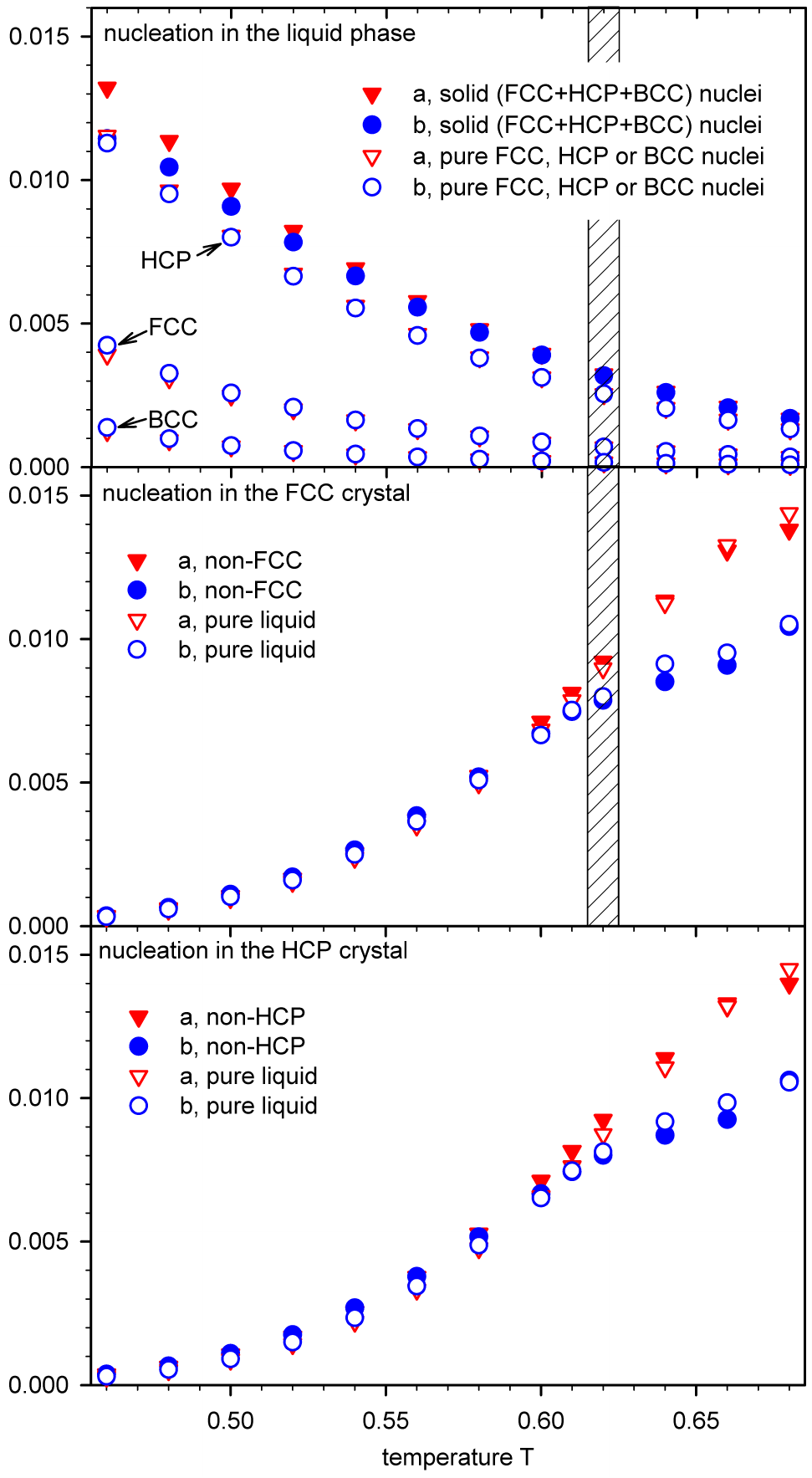}
\caption{\label{fig_a_b_vs_T_liq_FCC_HCP} Evolution with the temperature $T$ of the coefficients $a$ and $b$ giving the linear and quadratic terms of the development of $\psi(q)$ at small $q$ (Eq.~\ref{eq_psi} and \ref{eq_ab}). From top to bottom: nucleation in the liquid ($N=2000$), in the FCC crystal ($N=2048$) and in the HCP crystal ($N=2744$) respectively. In each case, several criteria have been considered to define the nuclei (given in the figures). The hatched region in the liquid and FCC panels corresponds to the temperature range where coexistence between liquid and FCC is likely to occur for a system of size $N \simeq 2000$ (see Fig~\ref{fig_T_coex}).
}
\end{figure}

As previously, these distributions have been characterized through the two coefficients $a$ and $b$ giving the linear and quadratic dependence of $\psi(q)$ (Eq.~\ref{eq_psi} and \ref{eq_ab}).
Figure~\ref{fig_a_b_vs_T_liq_FCC_HCP} gives their evolution with temperature.
The first point to be noticed is that $a$ and $b$ decrease when temperature increases for solid nucleation in liquid, while it it the converse for nucleation in the two solids (FCC and HCP).
It is also visible that the crystal structure (FCC versus HCP) has a negligible impact on $a$ and $b$ (absolute differences less than $10^{-4}$). Therefore, the crystal structure is shown to have a negligible influence on the first stages of nucleation.

Focusing on the FCC crystal, the data show that the two definitions of the nuclei (pure liquid vs non-FCC) give essentially the same results.
One can however notice few tiny differences, not imputable to statistical fluctuations since very similar features are observed for the HCP crystal. 
For nucleation in the liquid, in addition to the usual definition including all solids, the distributions associated with pure FCC, HCP or BCC nuclei were also analysed. The main point to be noticed is that the BCC nuclei are the less abundant, while HCP contribution is the largest, at odds with the fact that the FCC phase is the most stable. 

Focusing on solid nucleation in liquid, one finds $a=b$ for $T \geq 0.58$, while this equality fails below 0.56, the discrepancy increasing at lower temperatures. This means that the distribution $\phi(n)$ does not follow the Poisson law at low $T$, revealing a lack of independence between nucleation events.
Examination of nucleation in the two solids (FCC and HCP) shows a symmetric situation where $a=b$ below $T=0.58$ and departure from the Poisson law above $T=0.59$. Note that, in this case, the difference between $a$ and $b$ grows faster than for solid nucleation in the liquid.
As is clearly evidenced by the point at $T=0.61$, the difference between $a$ and $b$ appears strictly below the coexistence temperature $T_{sl}$ and is therefore not related to it.
Furthermore, one does not observe any discontinuity or change in slope while crossing the solid-liquid coexistence temperature: the change in stability of the phase (stable vs metastable) has no impact on the thermodynamic properties relevant for the first stages of nucleation.
As a matter of fact, the first stages are dominated by surface contributions, not influenced by the stability of the phase. On the other hand, for larger nuclei, where volume contributions play a role, the relative stability between the phases is expected to become important.  

It is also interesting to examine the data close to the freezing and melting temperatures.
One expects a rapid increase of the number of nuclei while approaching the limits of stability.
The coefficient $a$, giving the average number of nuclei, is thus expected to increase faster for the liquid around $T_f$ or for the solids around $T_m$.
This is qualitatively observed for solid nucleation in the liquid. On the other hand, nucleation in the two crystals exhibit an inflexion point below $T_m$, suggesting a saturation effect. Exploring higher temperatures, however, is made more difficult due to the rapid melting of the system.

\subsection{Distributions of nucleus size $p_a(s)$ and $p_l(s)$}
\subsubsection{Any vs largest nuclei}
\begin{figure}[]
\includegraphics[width=1.0\columnwidth]{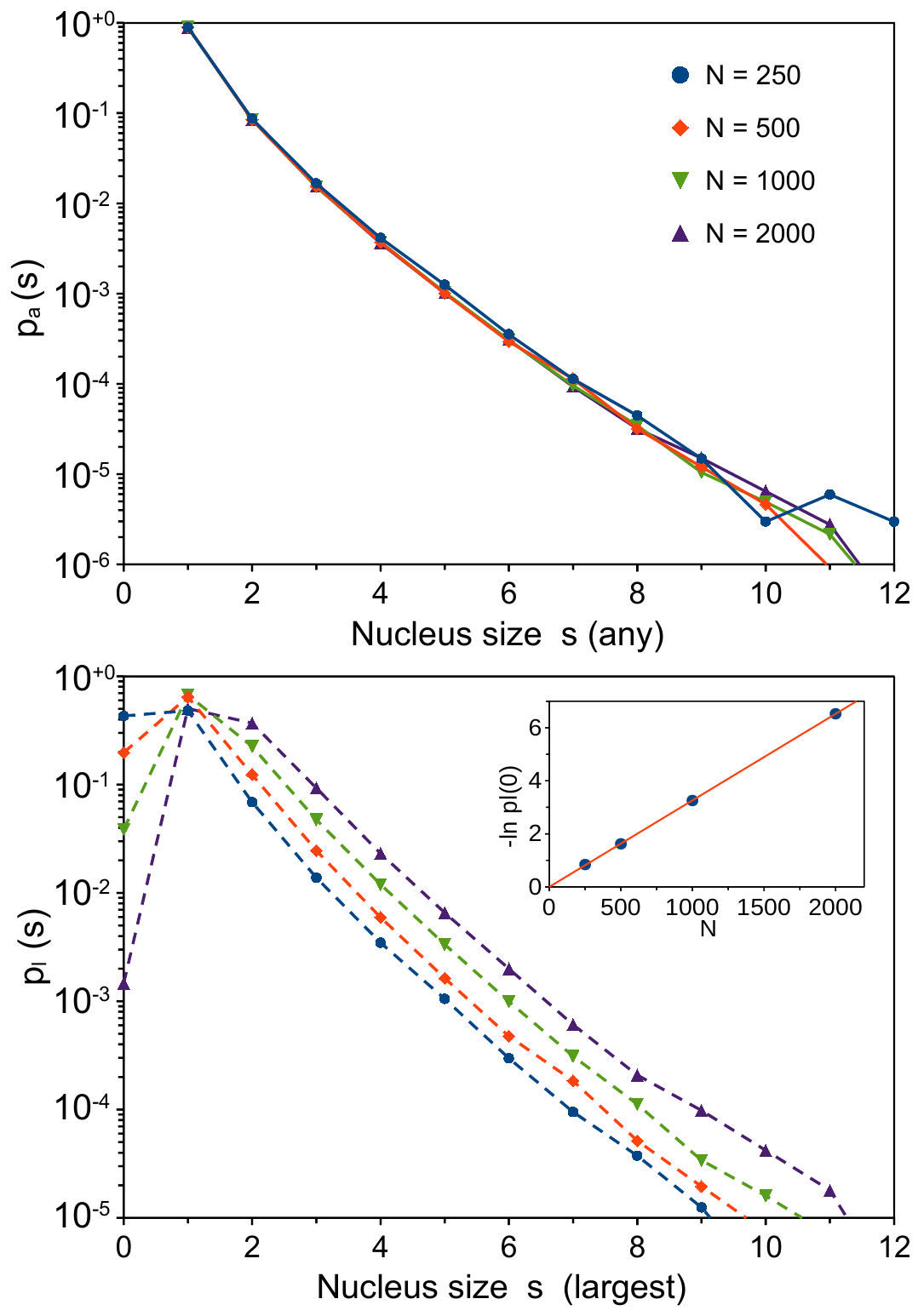}
\caption{\label{fig_any_vs_largest_freezing} Symbols: size distributions of any ($p_a(s)$, upper panel) and of the largest ($p_l(s)$, lower panel) solid (FCC+HCP+BCC) nuclei in the liquid at $P = 0$ and $T = 0.62$, for different system sizes $N$ given in the figure. Dotted lines (lower panel) are guides to the eye. Lines (upper panel): calculation of $p_a(s)$ from $p_l(s)$, as given by Eq~\ref{eq_pl_to_pa_discret}, following the same color code. Inset: $-\ln p_l(0)$ vs $N$. 
}
\end{figure}

The histograms giving the size distributions of any $p_a(s)$ or the largest $p_l(s)$ solid nuclei in the liquid at $T$ = 0.62 are given in Fig.~\ref{fig_any_vs_largest_freezing} for different system sizes ranging from $N$ = 250 to 4000.
As can be seen, $p_a(s)$ is independent of the system size, since all nuclei contribute equally to the distribution, independently of the others.
It is an intrinsic statistical property of the fluid, that depends only on intensive parameters like $T$ or $P$, relevant for nucleation processes, as given by Eq~\ref{eq_boltz}.

On the other hand, $p_l(s)$ is system size dependent. The larger the system, the lower the probability to observe a largest nucleus smaller than few atoms.
The most stringent situation is for $p_l(0)$ which corresponds to the probability not to observe nuclei in the system: it decreases exponentially with the system size (see inset).
Conversely, the probability to observe a largest nucleus with a large number of atoms increases with the system size.

\begin{figure}[]
\includegraphics[width=1.0\columnwidth]{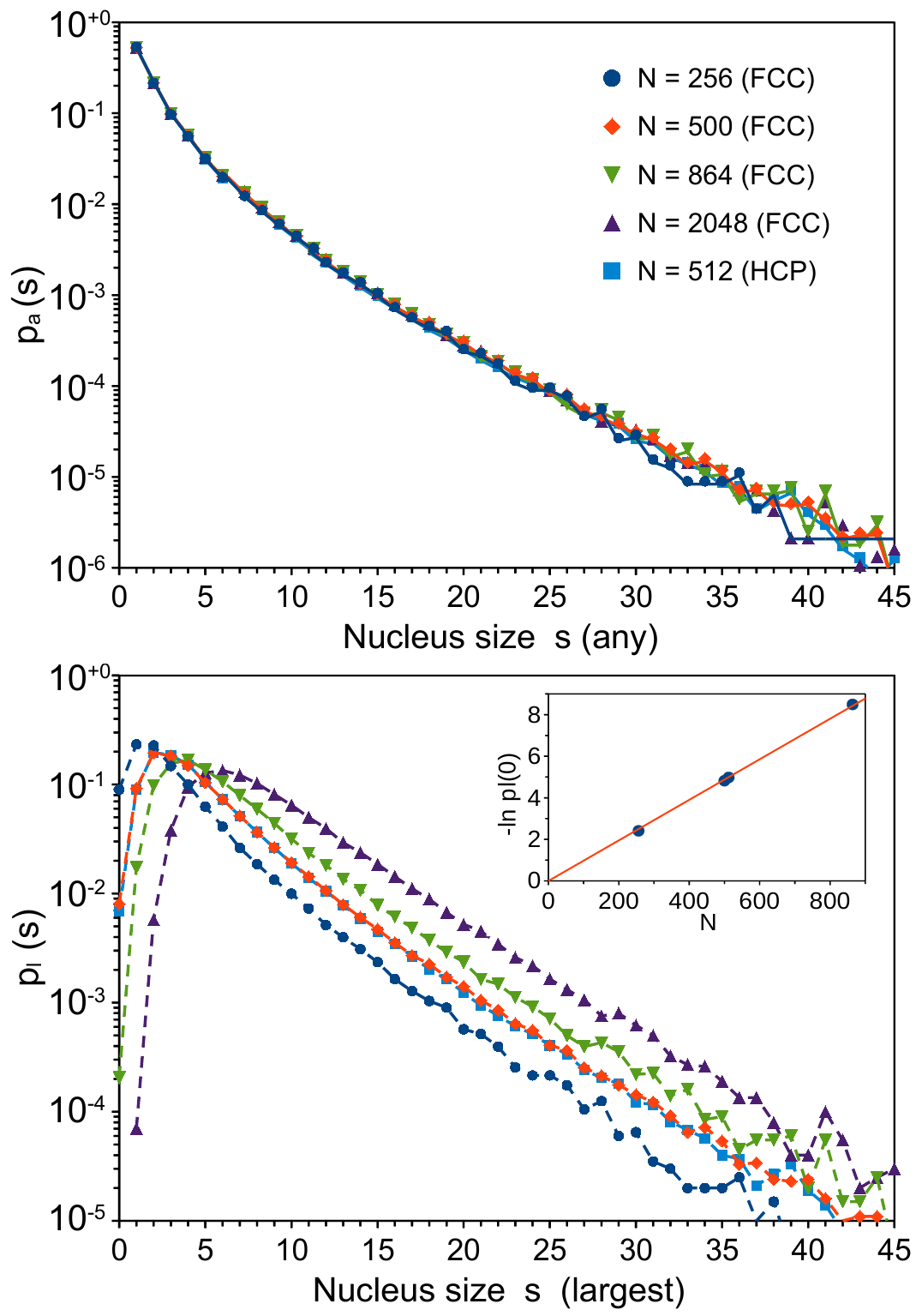}
\caption{\label{fig_any_vs_largest_melting} Symbols: size distributions of any ($p_a(s)$, upper panel) and of the largest ($p_l(s)$, lower panel) nuclei in the FCC and HCP crystals at $P = 0$ and $T = 0.62$, for different system sizes $N$ given in the figure. Dotted lines (lower panel) are guides to the eye. Lines (upper panel): calculation of $p_a(s)$ from $p_l(s)$, as given by Eq~\ref{eq_pl_to_pa_discret}, following the same color code. Inset: $-\ln p_l(0)$ vs $N$.
}
\end{figure}

The results of nucleation in the FCC crystal at the same temperature $T$ = 0.62 are given in Fig.~\ref{fig_any_vs_largest_melting}.
As can be seen, the same qualitative behavior is observed for the FCC crystal, i.e. a large system size dependence for $p_l(s)$ while $p_a(s)$ remains unchanged.
The only difference is quantitative: the probability to observe a given nucleus size is significantly lower for solid in liquid compared to the converse at the same coexistence temperature $T=0.62$. 

The HCP crystal exhibits quantitatively similar results to the FCC case. This is clear for $p_a(s)$ from Fig.~\ref{fig_any_vs_largest_melting}. For $p_l(s)$, since it depends on the system size, a direct comparison between FCC and HCP requires to choose the system sizes closest as possible: this is done for $N_{\rm HCP}=512$ which exhibits quantitatively the same distributions as for $N_{\rm FCC}=500$ (see Fig.~\ref{fig_any_vs_largest_melting}).

\subsubsection{Conversion of $p_l(s)$ into $p_a(s)$}
As mentioned in the Introduction, $p_a(s)$ is not accessible to simulations for large $s$, while $p_l(s)$ is calculable thanks to biased simulations for any size $s$. In Ref.~[\onlinecite{RN2954}] (Eq.~12) it was proposed the following algorithm to convert the discrete $p_l(s)$ into $p_a(s)$ which applies in our case since the distributions $\phi$ are close to Poisson law:
\begin{equation}
 p_a(s) = \frac{1}{\lambda} \left\{ \ln \Pi_l(s) - \ln \Pi_l(s-1) \right\} \text{ for } s \geq 1 \label{eq_pl_to_pa_discret} 
\end{equation}
where $\Pi_l(s) = \sum_{i=0}^s p_l(i)$ is the discrete cumulative distribution of $p_l(s)$ and $\lambda = -\ln p_l(0)$.
The results are given in the upper panels of Figs.~\ref{fig_any_vs_largest_freezing} and \ref{fig_any_vs_largest_melting} as solid lines. As can be seen, the results perfectly superimpose to the data $p_a(s)$ established by direct simulations, which validates the transformation.

\section{Discussion and conclusion}
\begin{figure}[]
\includegraphics[width=1.0\columnwidth]{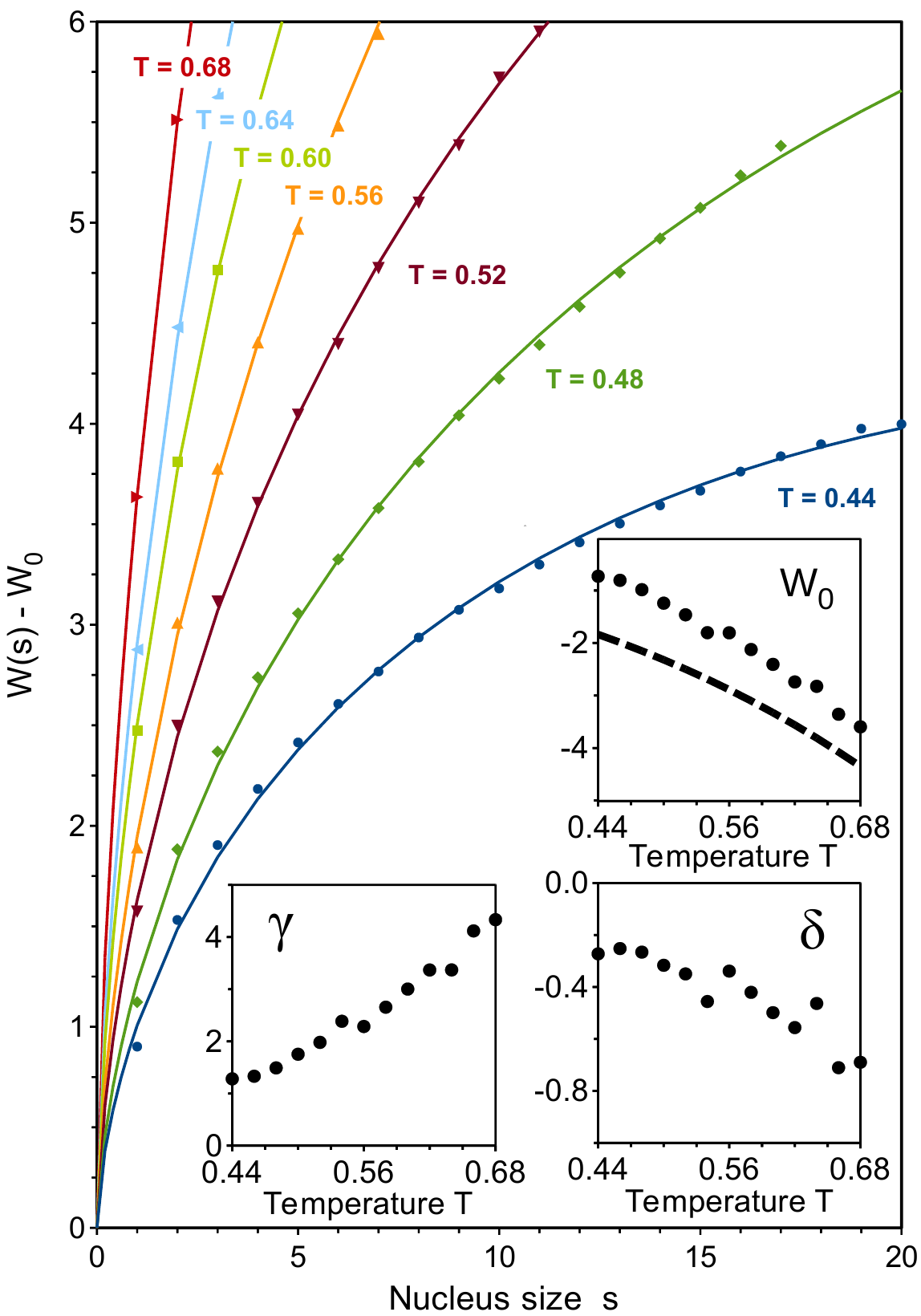}
\caption{\label{fig_free_energy_profiles_Liq} Main panel: Symbols: free energy profiles of solid nucleation in the liquid as given by Eq~\ref{eq_boltz} for different temperatures and for $N=2000$. Lines are best fits with Eq~\ref{eq_fit_free_E}; the evolution of the free parameters $\gamma$, $\delta$ and $W_0$ with $T$ is given as symbols in the insets; dashed line: see text and Eq~\ref{eq_w0}.  
}
\end{figure}

\begin{figure}[]
\includegraphics[width=1.0\columnwidth]{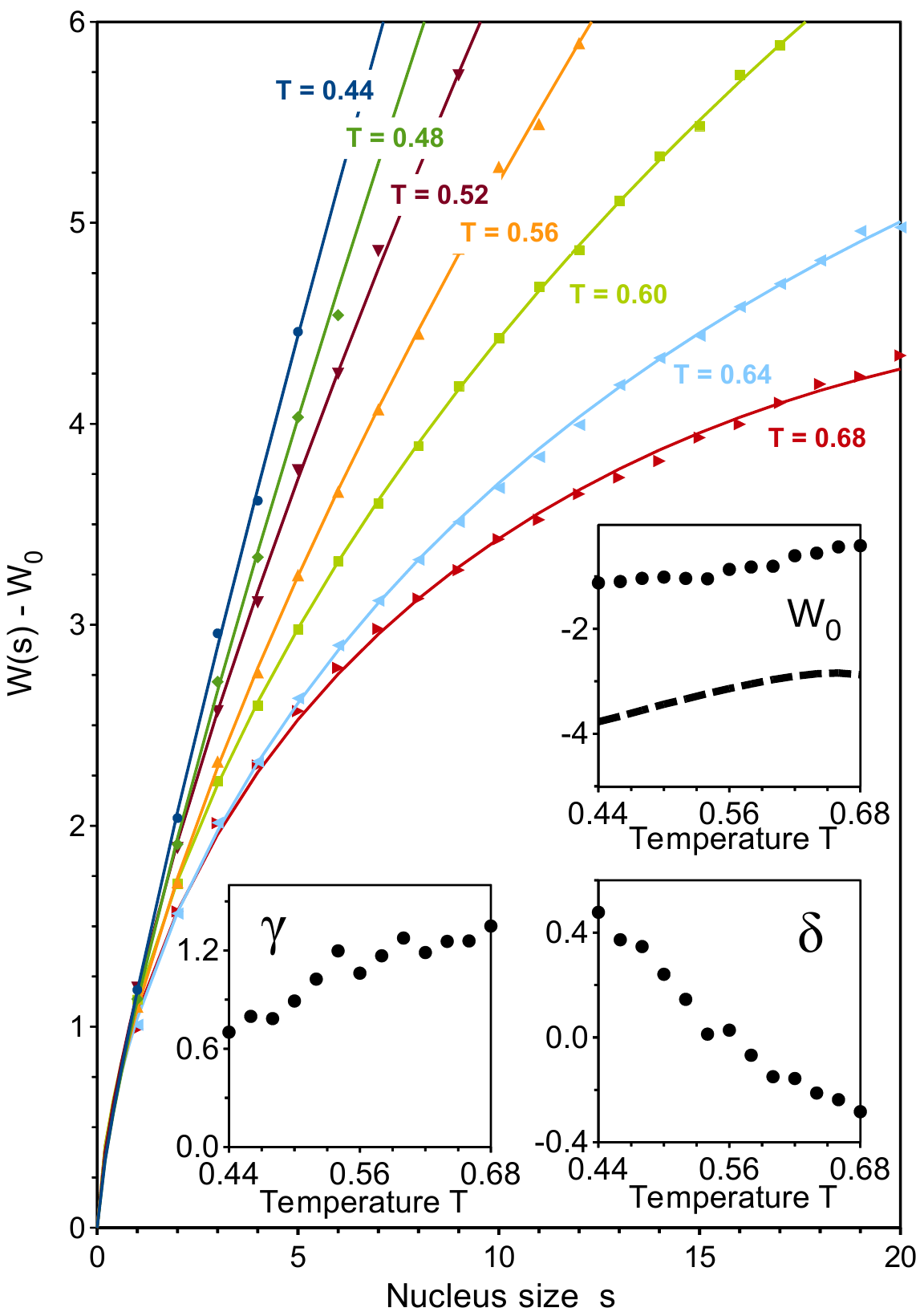}
\caption{\label{fig_free_energy_profiles_FCC} Main panel: Symbols: free energy profiles of nucleation in the FCC crystal as given by Eq~\ref{eq_boltz} for different temperatures and for $N=2048$. Lines are best fits with Eq~\ref{eq_fit_free_E}; the evolution of the free parameters $\gamma$, $\delta$ and $W_0$ with $T$ is given as symbols in the insets; dashed line: see text and Eq~\ref{eq_w0}.  
}
\end{figure}

The objective of this work was to validate Eq~\ref{eq_pl_to_pa_discret} in the case of the solid-liquid transition. In order to discuss nucleation, it is now required to calculate $p_l(s)$ for any nucleus size, in particular up to the top of the hill corresponding to the nucleation barrier. This requires lengthy biased simulations left for a future work.
It is however already possible to discuss the very first stages of nucleation. In particular, using Eq~\ref{eq_boltz}, the free energy profiles $W(s)$ can be calculated (up to an irrelevant constant to be determined below): they are shown in Figs.~\ref{fig_free_energy_profiles_Liq} and \ref{fig_free_energy_profiles_FCC} (symbols) for nucleation in the liquid and FCC crystal respectively. 
As can be seen, the free energy profiles are monotonously increasing with the nucleus size. Their evolution with temperature is such that the free energy profile is larger in the stable phase (liquid at high $T$ or FCC crystal at low $T$) and lower in the metastable phase (liquid at low $T$ or FCC crystal at high $T$).
In particular, for the most metastable cases, the free energy profiles suggest the existence of a maximum that was however out of accessible data. 

In the classical nucleation theory, the free energy profile $W(s)$ is that of a growing nucleus with a perfectly spherical shape.
The main contributions are homogeneous to surface and volume, corresponding respectively to the free energy of the interface and that of the difference between the old and the new phases.
We thus introduce the general equation:
\begin{equation}
 W(s) = W_0 + \gamma \hspace{1pt} s^{2/3} + \delta s \hspace{1pt}  \label{eq_fit_free_E} 
\end{equation}
where the free parameter $\gamma$ plays the role of an effective interface free energy and $\delta$ plays the role of a free energy difference between the phases.
Since it is not possible to define and count the number of nuclei of size $s=0$, the constant $W_0 = W(0)$ is left as a free parameter.
A best fit of the data with Eq~\ref{eq_fit_free_E} is shown in Figs.~\ref{fig_free_energy_profiles_Liq} and \ref{fig_free_energy_profiles_FCC} as lines, and the evolution of the free parameters with temperature is given in the insets.
The first point to be mentioned is that the parameters vary continuously with temperature, showing no discontinuity while crossing the coexistence temperature ($T$ = 0.62).
As expected, the surface parameter is always positive.
However, we observe two unexpected behaviors: (i) $\gamma$ takes different values depending on whether we are considering solid nucleation in the liquid or the converse, and (ii) it increases with increasing temperature.
For nucleation in the FCC crystal, the volume parameter $\delta$ roughly decreases from a positive value below the coexistence temperature to a negative value above that temperature: this behavior is in agreement with what is expected from a free energy difference between the bulk phases.
However, for solid nucleation in liquid, the behavior is at odds with expectations: the free energy difference is always negative.
It is however emphasized that we focused on the very first stages of nucleation where the volume contribution is small compared to the surface term.
Further calculations (at larger nucleus size) are required to discuss with a better accuracy the evolution of the parameter $\delta$ with $T$.

For ``rare'' clusters, the Gibbs free energy of formation is given by $\Delta G(s) = -kT \ln (n(s)/N) $ where $n(s)$ is the number of nuclei of size $s$ and $N$ is the total number of particles in the system.\cite{RN2911} 
According to our definition given by Eq~\ref{eq_boltz} and the fact that $n(s)/N = a_N p_a(s)$ where $a_N$ is defined by Eq~\ref{eq_ab}, one gets $W(s) = \Delta G(s) + kT \ln a_N$.
This leads to the expected value of the free parameter $W_0$:
\begin{equation}
 W_0 = kT \ln a_N \label{eq_w0}
\end{equation}
shown as a dashed line in the insets giving $W_0$ in Figs.~\ref{fig_free_energy_profiles_Liq} and \ref{fig_free_energy_profiles_FCC}.
As can be seen, $W_0$ almost follows the expected behavior, except for an overall constant. This difference probably originates in the fact that the ``rare event'' condition is clearly not fulfilled in our simulations, since several nuclei are frequently observed in our systems. This is intrinsic to the fact that we are considering tiny nuclei with large occurrence probability.

As a conclusion, the careful analysis of the number distributions of nuclei $\phi(n)$ in the stable and metastable phases exhibit no qualitative differences. 
The distributions follow essentially the expected Poisson law for independent nucleations, except in the case where the density of nuclei is so large that correlations may appear.
This is in agreement with the results of the finite size analysis showing no size dependence at low nuclei density.  
In this context, the relation (Eq~\ref{eq_pl_to_pa_discret}) between the size distributions of \textit{any} ($p_a$) or the \textit{largest} ($p_l$) nucleus is expected to hold, and this is what has been shown in the liquid as well as in the FCC and HCP crystal phases.
The main objective is now to use biased molecular simulations to establish $p_l(s)$ on the whole range of nucleus size $s$ in order to discus the evolution of the nucleation profiles with the temperature.

\begin{acknowledgments}
The authors are grateful to the CNRS interdisciplinary ``D\'efi Needs'' through its ``MiPor'' program (Project DARIUS), and to the financial support of Agence Nationale de la Recherche through the projects ANR-17-CE30-0002 (CavConf) and ANR-23-CE30-0028-03 (NanoCav).
\end{acknowledgments}

\section*{Data Availability Statement}
The data that support the findings of this study are available from the corresponding author upon reasonable request.

\section*{}\bibliography{solid_liquid}

\end{document}